\documentstyle[twoside,fleqn,espcrc2,epsfig]{article}

\newcommand{\AmS}{{\protect\the\textfont2
  A\kern-.1667em\lower.5ex\hbox{M}\kern-.125emS}}

\title{Fluxes of Atmospheric Neutrinos and Related Cosmic Rays}

\author{T.K. Gaisser\address{Bartol Research Institute,
        University of Delaware,\\
        Newark, DE 19716, USA}%
        \thanks{Research supported in part by the U.S. Department
            of Energy under Grant No. DE-FG02-91ER40626}}

\begin{document}

\begin{abstract}
The atmospheric neutrino beam simultaneously spans a range
of pathlengths from ten to ten thousand kilometers, which correspond
respectively to downward- and upward-going neutrinos.  As with
any neutrino oscillation experiment, also in this case the interpretation
of the data depends on a detailed knowledge of the neutrino beam.
The ingredients are the primary spectrum of cosmic-ray nucleons,
the geomagnetic fields in which the charged particles propagate
and the properties of interactions of hadrons in the atmosphere.
In this talk I review the status of calculations in light of the recent
evidence for neutrino oscillations from Super-Kamiokande \cite{evidence}.
\end{abstract}

\maketitle

\section{Introduction}
When cosmic ray protons and nuclei interact in the atmosphere, the secondary
cascades include neutrinos from decay of pions, muons and kaons.
Production of these neutrinos depends on the local zenith angle
because of the competition between decay and interaction of the
parent mesons in the tenuous atmosphere.   A simple geometric construction
\cite{Ayresetal} shows that a trajectory from below with nadir angle $\theta$
has the same zenith angle $\theta$ on the other side of the earth.
Therefore, since neutrinos with $E\ll 10^5$~GeV are virtually unattenuated by
the earth, the flux of atmospheric neutrinos would be up-down symmetric
in the absence of neutrino oscillations except to the extent that
the isotropy of the primary cosmic rays is distorted by the
geomagnetic field.   Variation of the neutrino flux
with azimuth is a consequence only of the geomagnetic field
(the ``East-West'' effect) even in the presence of oscillations
(because within a given band of zenith angle the
distributions of neutrino pathlengths and energies are independent
of azimuth).  This fact allows \cite{LSG} an important check
of the systematics of the Super-K analysis \cite{SuperK}, which I
discuss in \S 2.

Typical altitudes of production of the neutrinos are between
10 and 20 kilometers, so the distribution of neutrino pathlengths
ranges from 10~km for vertically downward neutrinos (neutrinos that
originate directly overhead) to $\sim 10^4$~km for upward neutrinos
from below.  The range
of neutrino energies for contained or partially contained events
is from sub-GeV to multi-Gev.  For neutrino-induced upward,
throughgoing muons it extends to $\sim 1000$~GeV.  Thus
the atmospheric neutrinos have a range of pathlength over
neutrino energy $1 < L_{km}\,/\,E_{GeV} < 10^5$.

The $\pi\rightarrow\mu\rightarrow e$ decay chain is the predominant
mode of production of atmospheric neutrinos in The sub-GeV to multi-GeV
range.  This leads to the basic prediction of
\begin{equation}
{\nu_e\,+\,\bar{\nu}_e\over\nu_\mu\,+\,\bar{\nu}_\mu}\; \sim \;{1\over 2}
\label{ratio}
\end{equation}
for $E_\nu\le 1$~GeV.  The ratio decreases as energy increases
because muons are increasingly likely to reach the surface before
decaying.  Comparison of decay length to energy-loss length in the
earth leads to the conclusion that virtually all muons that reach
the ground stop before decay (or capture) occurs.  Therefore muons
that reach the ground do not contribute the neutrinos with energy
high enough to contribute even to the sub-GeV sample ($p_e>100$~MeV/c).

In contrast with the expectation of Eq.~\ref{ratio},
several experiments \cite{IMB,Kam,Soudan,SuperK} find
\begin{equation}
R\;=\;{(\mu -like/e-like)_{data}\over (\mu -like/e-like)_{MC}}
\approx 0.65, \label{RRatio}
\end{equation}
which is equivalent to 
$(\nu_e+\bar{\nu}_e)/(\nu_\mu+\bar{\nu}_\mu)\approx 0.77.$
Ingredients that enter the denominator of the ratio of ratios in
Eq.~\ref{RRatio}
include the calculated neutrino flux, the cross sections for neutrinos
to interact in the quasi-elastic and various multi-prong channels and
the detection and reconstruction efficiencies in the detector.
The subject of this talk is the neutrino fluxes.

Three independent calculations \cite{BGS,HKHM,BN}
have been compared and analyzed \cite{GHKLMNS}
in order to identify and evaluate the sources of uncertainty in our knowledge
of the atmospheric neutrino flux.
Here I concentrate on comparison of the
calculations of Honda {\it et al.} \cite{HKHM,HKKM} with the
``Bartol fluxes''\cite{AGLS,GS}, because these two have been used by the
experimental groups \cite{SuperK,Soudan,MACRO} for analysis of their data.
The calculations of Refs.~\cite{AGLS}
and \cite{GS} are extensions of the calculations of Ref.~\cite{BGS}
respectively to high ($>3$~GeV) and to low ($<200$~MeV) energy.
In addition, a more realistic treatment of the geomagnetic cutoffs
is used \cite{LS}, which reduces the
calculated neutrino fluxes in the sub-GeV
range by about 10 per cent.  Honda {\it et al.} \cite{HKKM} also
extended their calculation to high energy.
At present the Bartol neutrino fluxes
and the Honda {\it et al.} fluxes in the GeV range agree within 5\% in
magnitude as well as ratio.  This level of agreement in magnitude is,
however, smaller than the systematic uncertainties, as I will discuss in
Sections 3 and 4.

In sections 3 and 4 I discuss the primary fluxes and the
treatment of hadronic interactions, both of which influence the
spectrum and shape of the neutrino spectra.
In the Super-K analysis, the overall normalization
is treated as a free parameter because of the large
uncertainty in the normalization of the primary spectrum.
There are recent
measurements of the primary spectrum that should in principle
allow one to reduce
this source of uncertainty.

As emphasized by Perkins \cite{Perkins},
muon fluxes high in the atmosphere are directly related to the
neutrino fluxes, being produced by the same primary spectra
and from the same interaction processes.  In \S 5 I
discuss how measurements of the flux of muons high in the atmosphere
are being used to check the overall normalization and shape of
the closely related neutrino flux.  In the conclusion I
list the various approximations
common to the present calculations and how they might be expected
to affect the results.

\section{Geomagnetic effects}

Propagation of a cosmic-ray nucleus through the geomagnetic
field depends only on its gyroradius and hence on the magnetic
rigidity,
\begin{equation}
R\;=\;A\times pc/(Ze)
\label{rigidity}\end{equation}
Here $A$ and $Z$ are the mass and charge
of a nucleus of momentum-per-nucleon $p$.
Low energy particles at low geomagnetic latitudes cannot reach
the atmosphere to produce secondaries.  Since energy per nucleon
is the important quantity for production of secondaries, nuclei
become relatively more important compared to protons at low
geomagnetic latitudes because of the factor $A/Z\approx 2$
in Eq. \ref{rigidity}.

Both neutrino flux calculations \cite{HKKM,AGLS} use
geomagnetic cutoffs obtained by the standard
method of backtracking antiprotons through the geomagnetic field
to determine the cutoffs for a particular location.  For example,
in the calculation of Ref. \cite{LS}, which is used in Ref. \cite{AGLS},
antiparticles are injected at 20 km altitude
on an outward trajectory.  If the trajectory reaches $30\,R_\oplus$
before it travels $500\,R_\oplus$ and
without intersecting the surface of the earth, then
it is assumed that positive particles of the same rigidity can
reach the atmosphere from that direction.

For the location of Super-K we have compared the cutoffs used in
Ref. \cite{HKKM} with those of Ref. \cite{LS} used for the calculation
of Refs. \cite{AGLS,GS}.  The cutoff maps are very similar, but with
some noticeable differences toward the east, where the cutoffs are
slightly higher in Ref. \cite{LS}.

At low geomagnetic latitudes such as Kamioka,
average cutoffs are higher locally (i.e. for cosmic rays entering the
atmosphere above the detector)
than for the opposite hemisphere (i.e. for the cosmic rays entering
the atmosphere on the other side of the earth, which give rise to
upward-going events).  The opposite is the case for a detector at
a high geomagnetic latitude, such as Soudan.  There the local cutoffs
are negligible in the sense that essentially all cosmic-rays from above with
sufficient energy to produce pions and contribute to the flux of neutrinos
can reach the atmosphere to interact.  Upward events originate from the
atmosphere over the entire hemisphere below each detector.  Since the average
over a full hemisphere is similar from any viewpoint, the
upward/downward ratio should be greater than one at Kamioka but
less than one at Soudan.

\begin{figure}[!htb]
\centerline{\epsfig{figure=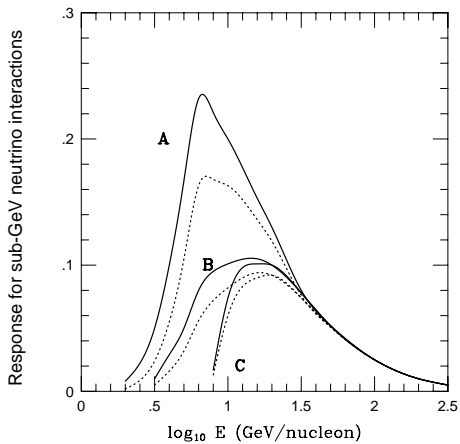,height=6.5cm}}
%\framebox[55mm]{\rule[-21mm]{0mm}{43mm}}
\caption{Response for sub-GeV muon-like events in Super-K.}
\label{fig:response}
\end{figure}

Fig. \ref{fig:response} illustrates the situation.
The pair of curves labelled (A)
shows the distribution of primary cosmic-ray energies that
would contribute to the sub-GeV signal in Super-Kamiokande if there
were no geomagnetic cutoff at all.  The solid curve is
for solar minimum and the dotted one for solar maximum.
The middle pair (B) is the corresponding
response from below, which would be similar if Super-K were moved to
Soudan.  The rightmost pair of curves (C) is the response for downward sub-GeV
events at Super-Kamiokande.  What is plotted is proportional to
the event rate per logarithmic interval of primary energy, so
in each case the area is proportional
to the signal.  Thus the upward/downward ratio at Super-K is $B/C > 1$.
If Super-K were located at Soudan, the ratio would instead be $B/A <1$.
(The method used to simulate ``sub-GeV'' events is described in
Ref.~\cite{LSG}.)

As neutrino energy increases the up-down asymmetry from the geomagnetic
effect diminishes.  For this reason, the Super-Kamiokande group have
emphasized the multi-GeV event sample in their search for neutrino
oscillations \cite{evidence,Kajita}.  On the other hand, the full
data set contributes to the evidence for oscillations.  Therefore it
is important to note \cite{LSG} that the geomagnetic effects themselves
provide a way of testing the integrity of the entire analysis chain
that is independent of whether or not there are oscillations.

At the low geomagnetic latitude of Kamioka there is a pronounced
east-west effect on the cosmic radiation.  Cutoffs are significantly
lower for positive particles from the west than from the east.  For example,
the trajectory of a 20 GeV antiproton injected toward the east from
above Super-K at $70^\circ$ from the zenith would be bent down by
the geomagnetic field and intersect the surface of the Earth, while
the same antiproton injected toward the west would escape from the geomagnetic
field.  In other words, the cutoff for protons
with zenith angle $70^\circ$ from the east at Super-K is $>20$~GeV.
For directions closer to the horizon
the cutoff  from the east approaches
$50$~GV.  In contrast, for directions above the horizon from the west
the cutoff is $5$~to~$10$~GV at Kamioka.

The excess of primary cosmic rays from the west at Kamioka
produces a corresponding
east-west asymmetry of the low-energy neutrino flux and
hence of the sub-GeV
event rate.  There is a much smaller, but still non-negligible asymmetry
for the multi-GeV event sample \cite{LSG}.
Since the east-west effect is an azimuthal asymmetry, it is independent
of oscillations; oscillation effects depend on neutrino pathlengths,
which vary with zenith angle but are independent of azimuth.

\begin{figure}[!htb]
\centerline{\epsfig{figure=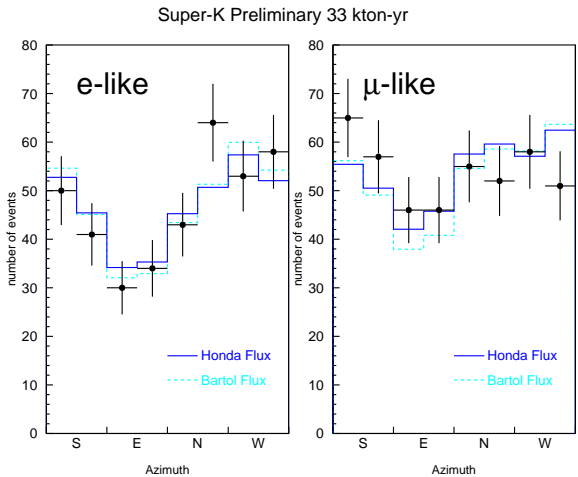,height=6.5cm}}
%\framebox[55mm]{\rule[-21mm]{0mm}{43mm}}
\caption{Azimuthal dependence: calculated and observed by Super-K.}
\label{fig:azimuthal}
\end{figure}

Figure \ref{fig:azimuthal}~\cite{azimuthal}
compares the azimuthal dependence of the Super-K data
($0.4<p_{lepton}<2.0$~GeV/c, single ring events
in 22.5 kton fiducial volume)
with expectation.  The solid line
uses the neutrino fluxes of Ref. \cite{HKKM} and the dashed line
the calculation of Ref. \cite{AGLS,GS}.
Although the fits are
equally good ($\chi^2/d.o.f.\approx 1$ for all four comparisons),
the geomagnetic effect is somewhat more pronounced with the
Bartol neutrino flux \cite{AGLS,GS} than with the flux of Honda {\it et al.}
\cite{HKKM}.  We have made some diagnostic tests to investigate
the source of this difference and a similar difference between
the two calculations that shows up in the zenith angle dependence
of sub-GeV events.  The difference arises in part from the
difference in cutoffs mentioned above, but also from the difference
in primary spectrum, as discussed below.

\section{Primary spectrum}

Both the normalization and the shape of the assumed primary
spectrum have important consequences for the calculation
of the neutrino fluxes.  The normalization propagates directly
through to the event rate.  The assumed spectral index affects the
shape of the neutrino energy spectrum in an obvious way,
but it also affects the angular dependence through its interaction
with the geomagnetic effects.  Thus, a softer spectrum will lead
to more pronounced geomagnetic effects because
a larger fraction of the event rate comes from lower energy
primaries, which are most affected by the geomagnetic field.

\begin{figure}[!htb]
\centerline{\epsfig{figure=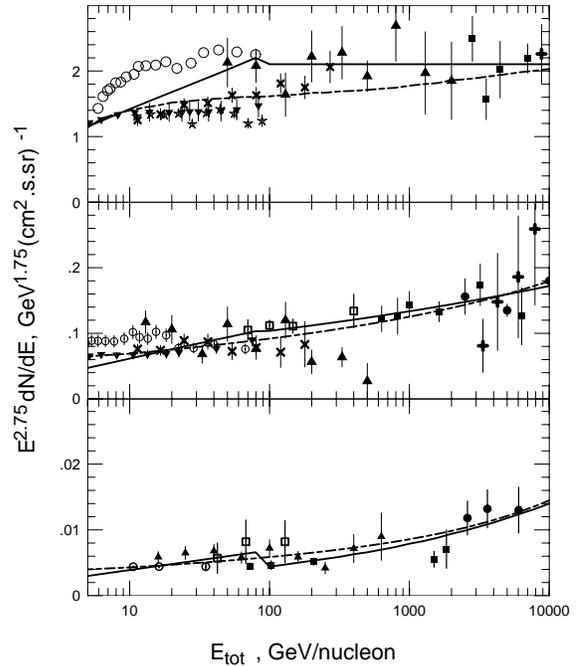,height=9cm}}
\caption{Summary of primary spectra.}
\label{fig:spectra}
\end{figure}

Fig. \ref{fig:spectra} is a summary of measurements of spectra of protons,
helium and the CNO group of nuclei, compared with the primary
spectra of Honda {\it et al.} \cite{HKKM} (solid lines) and the
spectra used in Ref. \cite{AGLS} (dashed lines).   From
Fig. \ref{fig:response}, it is apparent that  $5<E<50$~GeV/nucleon
is the most important region of the primary spectrum for sub-GeV
events.  The harder spectrum of Ref.~\cite{HKKM}
in this energy region, coupled with the geomagnetic field,
contributes significantly to the fact that the geomagnetic
effects are somewhat smaller in the neutrino spectra of Ref.~\cite{HKKM}
than in Ref.~\cite{AGLS}, as mentioned in the previous section.

A marked feature of the plot for hydrogen is the fact that
the data of Webber \cite{Webber} (shown by the open circles
in Fig.~\ref{fig:spectra}) are significantly higher
than those from the LEAP experiment \cite{LEAP} and other
more recent experiments in the same energy region.  The difference is
outside the error bars, indicating a systematic effect.
The most recent result is from the BESS detector \cite{Sat1},
and other recent experiments (Refs. \cite{IMAX,CAPRICE})
are included in the BESS compilation.  Generally (with the
possible exception of the measurement of Ref.~\cite{MASS}),
the interpretation of which is complicated by an unusually
strong level of solar modulation)
all the recent experiments are consistent with the LEAP results.
In fits to their data the Super-K group have treated the
overall normalization of their rates as a free parameter.
The primary spectra and their potential consequences for
interpretation of the data on neutrino interactions are discussed
more fully in talks given at the Satellite Symposium \cite{Sat1,Sat2,Sat3}.

\section{Yields}
It is important to note that the primary spectrum is not the
only source of uncertainty in the normalization and shape
of the energy spectrum of atmospheric neutrinos.  Uncertainties in
the yields of pions and kaons in interactions of hadrons
with nuclei of the atmosphere
are also important. Not all of phase space is covered
in accelerator measurements with nuclear targets.  For sub-GeV
events the
important range of beam energies is from a few GeV to several tens
of GeV (see Fig. \ref{fig:response}).  In this energy range
the atmospheric cascades are dominated by interactions
of nucleons, and nearly all neutrinos are from the
$\pi\rightarrow\mu\rightarrow e$ decay chain.

Existing measurements with beam energies around 20 GeV
and light nuclear targets measure pions only above 3 or 4
GeV \cite{expt1,expt2}, and there are significant differences
in how the lower energy pions are represented in the different
neutrino flux calculations, as discussed in Ref.~\cite{GHKLMNS}.
The pion multiplicities, and the momentum distributions as reflected by
the spectrum-weighted moments for pion production, are highest in
the calculation of \cite{BGS,AGLS,GS}.  This compensates to some
extent for the higher assumed proton spectrum of Ref.~\cite{HKKM}
with the result that the calculated neutrino fluxes (comparing
Refs.~\cite{AGLS} and \cite{HKKM}) differ by less than either
the primary spectrum or the yields.

Yields in a new calculation of Battistoni {\it et al.}~\cite{Italia}
are intermediate between those of Refs.~\cite{BGS} and \cite{HKHM}.

\section{Muons}
The same primary spectra and the same hadronic interactions determine
both muon and neutrino fluxes.  Therefore,
comparison with measurements of muons high in the atmosphere
offers a way to check directly the neutrino fluxes.
The most important range of altitudes for pion decay is
10 to 25 kilometers, which corresponds to atmospheric depths
of $\sim 20$ to $\sim 200$~g/cm$^2$.

Many of the same detectors referred to above in connection
with recent measurements of the primary spectrum have
also been used to measure the muon spectrum during
ascent through the atmosphere and on the ground.
The calculations of Refs.~\cite{AGLS} (and \cite{HKKM}) compare reasonably
well with the measurements of the MASS experiment \cite{Circella},
although there is a relative excess of muons below 1 GeV
in the calculation.  On the other hand, a recent comparison
between \cite{AGLS} and the HEAT measurements of muons \cite{Coutu}
showed better agreement in the shape of the spectrum but with
an overall excess of the calculation relative to the data of
as much as 50\% in some bins.

Measurements on the ground and at float altitude necessarily
have better statistics than data obtained during ascent.
It is possible that some of the discrepancies referred to above
could be a consequence of the short exposures during ascent.
Both for MASS \cite{Circella} and HEAT \cite{Coutu} there is
a tendency for better agreement between calculation and
measurement at float and at the ground than during ascent.
This is an active area with further potential for reducing
uncertainties in the flux of atmospheric neutrinos.

There are interesting possibilities with the muon measurements
for probing details of the calculations.   For one thing,
muon fluxes at float altitude reflect directly the primary
spectrum and the properties of pion production in single
nucleon-nitrogen interactions with no intervening cascading.

A more interesting possibility arises from the fact that in
some cases the same detector has been exposed at different
locations with different geomagnetic cutoffs.   The MASS
experiment has been flown both in Northern Canada (essentially
no cutoff) and from Ft. Sumner, NM where the vertical cutoff
is $\approx 5$~GV.  The BESS detector has measured the muon
charge ratio on the ground in northern Canada and in Japan.
The low-energy behavior of the ratio is quite different in
the two locations.  Below $\sim1$~GeV the $\mu^+/\mu^-$ ratio decreases
toward 1 in Japan, which can be understood as a consequence
of the high local geomagnetic cutoff. There are two effects:\\
First, with a high cutoff, heavy primaries are relatively more
important because they have a higher rigidity for a given energy per nucleon.
Protons produce more positive than negative pions (and hence more $\mu^+$)
and {\it vice versa} for neutrons.  Thus, enhancing the contribution
from nuclei, which carry the neutrons,
suppresses the muon charge ratio slightly.
Secondly, vertical $\mu^+$ at the ground have followed trajectories
from slightly east of vertical (where the cutoffs are higher), whereas
vertical $\mu^-$ will have come slightly from the west where more
of the primary spectrum reaches the atmosphere to produce secondaries.
This enhances the negative relative to the positive muons and hence
reduces the $\mu^+/\mu^-$ ratio preferentially at low energy
where the bending is more significant.

\section{Conclusion}

Present calculations \cite{HKKM,AGLS} include several approximations:
\begin{itemize}
\item They are one-dimensional; i.e. all neutrinos are assumed
to follow the direction of the primary nucleon that produced them.
This approximation has two effects:
\begin{enumerate}
\item There should be some loss of particles that are produced
at large angle.  Given the momentum involved, as compared with
the typical transverse momentum of produced pions, it is straightforward
to check that this effect should be small for neutrino events in Super-K.
\item Bending of charged particles in the atmosphere is not followed.
This is perhaps the most important effect to check \cite{Goldhaber} 
because it is systematic.   As explained above, the vertical
muon charge ratio is reduced when the cutoffs are high.  There is
a corresponding decrease in the $\nu_e/\bar{\nu}_e$ ratio
(and an increase in that part of the $\nu_\mu/\bar{\nu}_\mu$ ratio
that comes from muon decay).  Because $\sigma_\nu > \sigma_{\bar{\nu}}$,
the calculated ratio of electron-like to muon-like events will
decrease with respect to the one-dimensional calculation.  This
correction will therefore make the anomaly of Eq. (\ref{RRatio}) 
somewhat more pronounced.
\end{enumerate}
\item The superposition model has been used in Ref.~\cite{AGLS}
for interactions of nuclei.
Within the framework of a standard multiple scattering picture,
this approximation can be shown to give a good account of the
distribution of first interactions of each nucleon.  It will,
however, lead to some overestimate of the multiplicity of
pions in the target fragmentation region.
\item The cascades are propagated to sea-level all over the
globe.  In particular, the exact terrain over the detector
(i.e. the mountain in the case of Super-K) has been neglected \cite{Spiro}.
This is negligible for muon neutrinos from pions, which decay
high in the atmosphere.  From the pathlength distributions
of Ref.~\cite{pathlength} it is possible to estimate the size of
the effect of this approximation.  To take an extreme case
of a 4 km overburden ($\sim$NUSEX), the neutrinos from
muon decay overhead are overestimated by about 10\%, leading
to a $\sim5$\% overestimate of the calculated
$(\nu_e+\bar{\nu}_e)/(\nu_\mu+\bar{\nu}_\mu)$ ratio.
\item The calculations are based on parametrizations of data in limited
regions of phase space.  Interpolations and extrapolations introduce
some level of uncertainty.  The yields of Ref.~\cite{AGLS} are
at the high end of a spectrum, with \cite{BN} the lowest and
\cite{HKKM} and \cite{Italia} in between.
\end{itemize}
At least two groups are embarking on three-dimensional calculations.
The Italian group \cite{Italia} has published a short account
of their plan with a comparison of their one-dimensional results
with those shown in Ref.~\cite{GHKLMNS}.  They use 
FLUKA~\cite{FLUKA} with various hadronic interaction models at different
energies.
The authors of Refs.~\cite{BGS} and \cite{AGLS}, together with Coutu,
are also pursuing this goal.  Although effects are generally
expected to be small, in view of the importance of the experimental
results, a greater level of detail in the calculations is warranted.

\vspace{.3cm}

\noindent
{\bf Acknowledgements}.  I am grateful to Ed Kearns for providing
me with Fig. 2~\cite{azimuthal} and to Todor Stanev for reading
the manuscript and for collaboration on this work.  I thank
M. Goldhaber and M. Spiro for useful conversations, and M. Honda
for very helpful exchanges of information about the calculations
of Refs.~\cite{HKHM,HKKM}.


\begin{thebibliography}{9}
\bibitem{Goldhaber} Maurice Goldhaber has emphasized this point (private
communication).
\end{thebibliography}
\end{document}